# Optimal Design of Shaped Pupil Lyot Coronagraph for China's Large Segmented Telescope

**Kaining Yang[1*], He Sun[1,2,3*], Bin B. Ren[4*], Fabo Feng[5], Yiyun Wu[5], Jinxin Liu[1], Linbo Cao[1], Ruobing Dong[6,7], Feng Long[7], Wei Wang[8]**

[1] *College of Future Technology, Peking University, Beijing* 100871, *China*;
[2] *Academy for Advanced Interdisciplinary Studies, Peking University, Beijing 100871, China*;
[3] *National Biomedical Imaging Center, Peking University, Beijing 100871, China*;
[4] *Department of Astronomy, Xiamen University, Xiamen, Fujian* 361005, *China*;
[5] *State Key Laboratory of Dark Matter Physics, Tsung-Dao Lee Institute & School of Physics and Astronomy, Shanghai Jiao Tong University, Shanghai 201210, China*;
[6] *Department of Astronomy, School of Physics, Peking University, Beijing 100871, China*;
[7] *The Kavli Institute for Astronomy and Astrophysics, Peking University, Beijing 100871, China*;
[8] *CAS Key Laboratory of Optical Astronomy, National Astronomical Observatories, Chinese Academy of Sciences, Beijing 100101, China*

*Correspondence: knyang25@stu.pku.edu.cn； hesun@pku.edu.cn; rbb@xmu.edu.cn

**Abstract:** China's next-generation large telescopes, such as the Expandable Aperture Segmented Telescope (EAST) and the Large Optical/Infrared Telescope (LOT), are being planned as flagship facilities for high-angular-resolution astronomy, including the direct imaging and characterization of exoplanetary systems (e.g., exoplanets, circumstellar disks). However, their segmented, centrally obscured apertures present significant challenges for coronagraph design: diffraction from segmented gaps and spiders severely limit contrast, particularly at small inner working angles (IWAs). To address this, we investigate an optimal Shaped Pupil Lyot Coronagraph (SPLC) architecture tailored for these telescopes, which combines a binary shaped pupil apodizer, a focal-plane occulting mask, and a Lyot stop to suppress stellar glare and enhance contrast. We further augment the coronagraph design with a two-deformable-mirror (2-DM) adaptive optics (AO) system to correct atmospheric turbulence and residual aberrations. End-to-end simulations based on site parameters for Lenghu in Qinghai Province and assuming idealized extreme AO (ExAO) correction demonstrate that the optimized 2DM-SPLC designs can achieve raw contrasts of order $10^{-6}$-$10^{-7}$ in the V, H, and K bands, providing an upper-bound estimate of the system performance. The simulated performance highlights the advantages of the SPLC in mitigating diffraction from segmented and centrally obscured pupils while maintaining a small IWA and practical throughput, indicating that the 2DM-SPLC is a promising coronagraphic architecture for high-contrast exoplanet imaging on China's future large segmented telescopes.

**Keywords:** Exoplanet detection; Shaped pupil Lyot coronagraph; Convex optimization; Adaptive optics; deformable mirror

## 1. INTRODUCTION

China is actively advancing its efforts to construct large optical telescopes and establish ground-based, large-aperture, general-purpose optical platforms. For example, Expandable Aperture Segmented Telescope (EAST) will be built with the primary mirror expanding from 6 m in the initial stage to 8 m in the second [1]. In addition, the Large Optical/Infrared Telescope (LOT) has been proposed, featuring a 14.5 m primary mirror composed of 120 hexagonal segments. This will be the largest general-purpose optical telescope ever built in China [2,3]. The development of EAST and LOT will complement existing large optical/infrared facilities worldwide and help close the current gap in general-purpose large-aperture optical telescopes in China. These facilities will also provide powerful ground-based follow-up capabilities for space observatories such as the James Webb Space Telescope (JWST) [4], the upcoming China Space Station Telescope (CSST) [5], Roman Space Telescope (RST) [6], and even the future missions including Tianlin [7,8] and Habitable Worlds Observatory (HWO) [9], by enabling high-sensitivity spectroscopy and high–angular-resolution imaging of selected targets. With their large collecting areas and advanced instrumentation, EAST and LOT will be particularly well suited to faint and high-resolution observations, including the direct imaging and characterization of exoplanets, detailed studies of protoplanetary and debris disk structures, quasar host galaxies, and investigations of compact objects and a wide range of transient phenomena.

**Table 1. Coronagraphic performance of large ground-based telescopes**

| Telescope | Aperture | Coronagraph | Band | Design IWA | Design Contrast | Raw Contrast (separation) | Post-processing Contrast (separation) |
|---|---|---|---|---|---|---|---|
| Magellan [10,11] | 6.5 m | MagAO-X | V (Hα) | 0.02″ | $1 \times 10^{-4}$ | / | $1 \times 10^{-5}$ (0.10″) |
| VLT[12,13] | 8.2 m | SPHERE | H | 0.08″ | $1 \times 10^{-5}$ | $8 \times 10^{-5}$ (0.50″) | $1 \times 10^{-6}$ (0.50″) |
| Gemini[14-16] | 8.1 m | GPI | H | 0.13″ | $1 \times 10^{-7}$ | $1 \times 10^{-4}$ (0.40″) | $1 \times 10^{-5}$ (0.40″) |
| Subaru[17,18] | 8.2 m | SCExAO | H | 0.08″ | $1 \times 10^{-5}$ | $8 \times 10^{-4}$ (0.20″) | $8 \times 10^{-5}$ (0.30″) |
| Keck[19] | 10 m | KPIC | K | 0.04″ | $1 \times 10^{-6}$ | / | $9 \times 10^{-6}$(0.42″) |

Recent advancements in extreme adaptive optics (ExAO) and coronagraphic instruments have significantly improved the high-contrast imaging capabilities of ground-based telescopes [20], as summarized in Table 1. Representative systems have demonstrated this progress across different wavelength ranges. In the visible, Magellan Adaptive Optics (MagAO) obtained the first ground-based visible image of β Pic b [21]. In the near-infrared, Gemini Planet Imager (GPI) and Spectro-Polarimetric High-contrast Exoplanet Research (SPHERE) have achieved smaller inner working angles and more stable dark holes, enabling both representative planet detections and systematic surveys. GPI discovered and characterized 51 Eri b [22] and conducted the Gemini Planet Imager Exoplanet Survey (GPIES) [23], while SPHERE detected planets such as HIP 65426 b and PDS 70 b/c [24-27] and carried out the SPHERE Infrared Survey for Exoplanets (SHINE) to constrain the demographics of wide-orbit giant planets [12,28]. Subaru Coronagraphic Extreme Adaptive Optics (SCExAO) and Keck Planet Imager and Characterizer (KPIC) further extend high-contrast imaging toward close-in companions and high-dispersion spectroscopic characterization, including molecular, rotational, and kinematic measurements of planetary systems such as HR 8799 [17,29-34]. Next-generation instruments, including MagAO-X and GPI 2.0, are being developed to further improve IWA, dark-hole stability, and observing flexibility [35-39]. China's Portable Adaptive Optics (PAO) has also demonstrated direct exoplanet imaging on a 3.5 m telescope by recovering κ And b [40]. Thirty-meter-class extremely large segmented telescopes are also expected to host high-contrast instruments for exoplanet imaging, including Giant MagAO-X (GMagAO-X) on Giant Magellan Telescope (GMT) [41], Exoplanet Imaging Camera and Spectrograph (EPICS) on Extremely Large Telescope (ELT) [42], Planetary Systems Imager (PSI) on Thirty Meter Telescope (TMT) [43]. Spatial-light-modulator-based coronagraphs have also been explored to mitigate segment-gap diffraction [44]. These developments highlight the essential role of advanced coronagraphs in exoplanet and circumstellar disk studies [45]. Upcoming space-based telescopes, including CSST and RST, will also carry dedicated coronagraphic instruments, namely Cool Planet Imaging Coronagraph (CPI-C) and Roman Coronagraph Instrument (CGI) [46-48].

A coronagraph improves the detectability of faint circumstellar targets by suppressing on-axis stellar glare with focal- and/or pupil-plane masks [20,45,49]. The classical Lyot coronagraph uses a focal-plane occulting mask and a Lyot stop to block the stellar core and remove diffracted light [50]. Building on this concept, numerous variants have been developed to improve performance. Table 1 summarizes key parameters of different coronagraph designs, including the observation band, inner working angle (IWA), raw contrast, and post-processing contrast. Here, "raw contrast" denotes the suppression from the coronagraph and wavefront control, whereas "post-processing contrast" includes additional gains from point-spread-function (PSF) subtraction and other differential-imaging techniques. Amplitude-type coronagraphs, including the band-limited coronagraph (BLC) and the apodized pupil Lyot coronagraph (APLC) [51-53], use pupil-plane apodization, such as microdot/halftone or step-/transmission-filter apodizers [54,55], together with a focal-plane mask to form dark holes, and have been widely adopted in the GPI [14] and SPHERE [12]. Phase-type coronagraphs, including the four-quadrant phase mask (4QPM), vector vortex coronagraph (VVC), and phase-induced amplitude apodization (PIAA) [56-59], can provide smaller IWAs and high throughput by modifying the focal-plane phase or remapping the pupil, but they are generally more sensitive to pointing errors and finite stellar angular diameter. These designs have been incorporated into several instruments such as SPHERE, SCExAO, and MagAO-X [12,17,35,60]. Other architectures, such as the high-dispersion coronagraph used by the Keck Planet Imager and Characterizer (KPIC), combine coronagraph with high-resolution spectroscopy for complementary exoplanet characterization [31].

A further concept is pupil shaping, in which the telescope pupil is engineered to create prescribed dark holes in the image plane [61]. A representative implementation is the Shaped Pupil Lyot Coronagraph (SPLC) [62], which replaces the grayscale apodizer of a traditional APLC with an optimized binary pupil apodizer (Fig. 1(a)). This design freedom allows the SPLC to maximize throughput within a specified dark hole, and provides higher energy efficiency than grayscale apodizers. Standard SPLC designs typically deliver an IWA of ~3–4 $\lambda/D$, while joint optimization of the pupil apodizer and Lyot stop can reduce it to ~2 $\lambda/D$ [62]. These advantages have motivated shaped-pupil architectures for the Roman Space Telescope Coronagraph Instrument, on-sky demonstrations with the SCExAO shaped-pupil coronagraph [63], and SPLC-like binary apodizers in the GPI 2.0 upgrade [37-38]. For future large segmented telescopes, SPLCs are particularly attractive because they can be co-designed to mitigate diffraction from segment gaps and secondary-mirror supports while forming high-contrast dark holes with adaptive optics. Therefore, deploying SPLCs on forthcoming large telescopes could enhance exoplanet imaging capabilities and help exploit the scientific return of large apertures.

The final contrast on ground-based telescopes depends not only on coronagraph design but also on the control of atmospheric and instrumental wavefront errors. Atmospheric turbulence imposes the seeing limit, while adaptive optics (AO) systems use wavefront sensors and deformable mirrors (DMs) to correct phase distortions in near real time and recover near-diffraction-limited resolution [64-68]. Depending on the science goal, AO systems range from single-conjugate narrow-field correction to multi-conjugate wide-field correction [69-74], whereas extreme adaptive optics (ExAO) is optimized for high-contrast exoplanet imaging [12,17,75,76]. However, even with AO correction, achieving contrasts of $10^{-6}$ or better remains challenging because residual static and quasi-static aberrations from telescope optics, thermal drifts, and non-common-path errors generate speckles after coronagraphic suppression, potentially overwhelming faint planetary signals [77]. To address this challenge, focal plane sensing and control (FPSC), an ExAO technique, uses the science camera to estimate and suppress wavefront errors directly in the coronagraphic image plane. By applying known phase modulations in the coronagraphic focal plane, FPSC measures the residual stellar electric field and drives deformable mirrors to form high-contrast dark holes around the star [78-88]. Representative algorithms include speckle nulling [78] and electric field conjugation (EFC) [79], which have been demonstrated on high-contrast testbeds and instruments for suppressing quasi-static speckles and improving contrast. Together with coronagraphs, FPSC enables ground-based telescopes to approach the performance required for direct exoplanet imaging and characterization.

In this paper, we develop two-DM SPLC (2DM-SPLC) systems for the EAST and LOT. The preloaded DM surfaces are optimized with the SPLC to generate and stabilize high-contrast dark holes for segmented apertures. We evaluate the system performance under wavefront errors, quantify the resulting changes in inner working angle, throughput, and contrast, and assess the recovery after FPSC correction. Finally, we analyze the capability of the proposed systems for detecting representative exoplanets.

## 2. METHOD

### 2.1 Optical System of 2DM-SPLC

The 2DM-SPLC architecture consists of an SPLC module and a two-DM wavefront modulation module. The SPLC includes a shaped pupil (SP), a focal plane mask (FPM), and a Lyot stop (LS): the SP apodizes the telescope pupil to form prescribed dark holes, the FPM blocks the stellar core, and the LS removes diffracted light. In the two-DM module, DM2 is placed at the pupil plane, while DM1 is located upstream at a plane conjugated to neither the pupil nor the focal plane. This DM configuration enables joint phase and amplitude modulation of the complex wavefront through propagation between the two DMs. With AO algorithms, the DMs can both pre-shape the wavefront for dark-hole generation and correct time-varying aberrations during observations.

The optical layout of the 2DM-SPLC is depicted in Fig. 1(a). Light from the telescope passes through DM1 and DM2, and then enters a 4-f relay containing the SP, which reshapes the pupil illumination, concentrates energy along the y-axis, and creates dark holes along the x-axis. After propagation to the focal plane, the FPM suppresses the bright stellar core while transmitting the dark-hole region. The field is then re-imaged onto the LS, which blocks diffracted light from the telescope aperture and masks, before the final coronagraphic image is recorded by the science camera.

### 2.2. Optimal Design of SPLC

**Telescope pupils.** The SPLC design is strongly constrained by the telescope aperture geometry. Both EAST and LOT adopt segmented primary mirrors made of identical hexagonal segments with a diagonal of 1.44 m. EAST includes two configurations: the initial 6 m phase with 18 segments and the expanded 8 m phase with 36 segments, while LOT consists of 120 segments with an effective diameter of approximately 14.5 m. These centrally obscured and segmented pupils introduce strong diffraction features that must be suppressed by the SPLC masks.

**Optical Modeling.** Light propagation within the SPLC is modeled using Fourier optics. The complex field $E_1$ in a pupil plane is transformed to the field $E_2$ in the subsequent focal plane via a discrete Fourier transform (DFT):

$$E_2(u,v) = \frac{e^{i\frac{2\pi f}{\lambda}}}{i\lambda f}\sum_{x=-\frac{1}{2}x_1}^{\frac{1}{2}x_1}\sum_{y=-\frac{1}{2}y_1}^{\frac{1}{2}y_1} E_1(x,y)e^{-i\frac{2\pi}{\lambda f}(ux+vy)}\Delta x_1\Delta y_1, \quad (1)$$

where $\lambda$ is the wavelength, $f$ is the focal length of the lens, $\Delta x_1$ and $\Delta y_1$ are the physical sampling intervals of $E_1$. Since the telescope aperture and masks are symmetric about both the $x$- and $y$-axis, the complex DFT can be replaced by a real-valued cosine transform over one quadrant, reducing the computational cost by approximately four times. The optical Fourier transform through a lens is denoted as $F_{\mathrm{opt}}[\cdot]$,

$$E_2(u,v) = 4\left|\frac{e^{i\frac{2\pi f}{\lambda}}}{i\lambda f}\right|\sum_{x=-\frac{1}{2}x_1}^{0}\sum_{y=-\frac{1}{2}y_1}^{0} E_1(x,y)\cos\left[\frac{2\pi}{\lambda f}ux\right]\cos\left[\frac{2\pi}{\lambda f}vy\right]\Delta x_1\Delta y_1 = F_{\mathrm{opt}}[E_1(x,y)]. \quad (2)$$

Consequently, the SPLC transfer from the entrance pupil to the final coronagraphic focal plane can then be written as

$$E_{\mathrm{C}} = F_{\mathrm{opt}}\left[F_{\mathrm{opt}}^{-1}\left[F_{\mathrm{opt}}[E_{\mathrm{in}}\cdot M_{\mathrm{AP}}\cdot M_{\mathrm{SP}}]\cdot M_{\mathrm{FPM}}\right]\cdot M_{\mathrm{LS}}\right] = T_C[E_{\mathrm{in}}], \quad (3)$$

where $E_{\mathrm{in}}$ is the incoming field, $T_C[\cdot]$ denotes the SPLC transfer model, $M_{\mathrm{AP}}$ is the telescope aperture, $M_{\mathrm{SP}}$, $M_{\mathrm{FPM}}$, $M_{\mathrm{LS}}$ are the binary quadrant masks of the SP, FPM, and LS, respectively. The FPM and LS are binary, while the SP is optimized using a continuous relaxation and yields a nearly binary solution. To reduce the computational cost, the field is evaluated only on the support of the corresponding mask at each plane, resulting in a sparse model suitable for large-scale optimization.

**Contrast metric.** The raw coronagraphic contrast is defined as the mean intensity in the dark-hole region, normalized by the peak intensity of the non-coronagraphic PSF. Let $E_{\mathrm{DH}}$ denote the electric field in the dark hole and $E_{\mathrm{Norm}}$ the on-axis PSF amplitude without the FPM; the mean contrast $C$ is given by:

$$C = \frac{\langle|E_{\mathrm{DH}}|^2\rangle}{|E_{\mathrm{Norm}}|^2},$$
$$E_{\mathrm{Norm}} = 4\left|\frac{e^{i\frac{2\pi f}{\lambda}}}{i\lambda f}\right|\mathrm{sum}\left[M_{\mathrm{SP}}|_{(x,y)\in M_{\mathrm{LS}}}\right]\cdot\sqrt{\Delta x_1\Delta y_1}\cdot\sqrt{\Delta x_3\Delta y_3}. \quad (4)$$

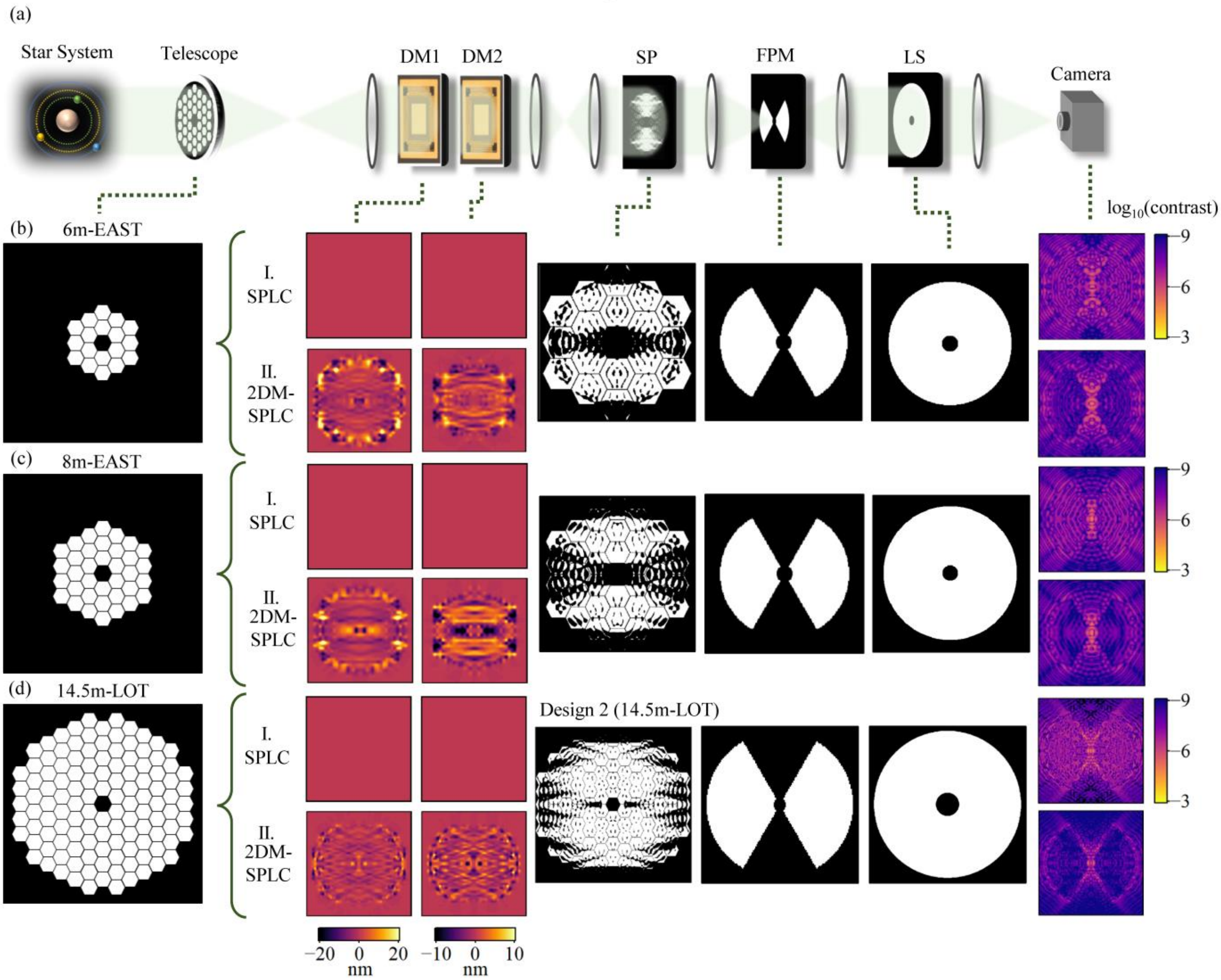

**Fig. 1. Optical setup and the design of the SPLC and 2DM-SPLC.** (a) Schematic layout of the optical components in the 2DM-SPLC system. DM: deformable mirror; SP: shaped pupil; FPM: focal plane mask; LS: Lyot stop. The telescope is a large-aperture reflective system. The SPLC relay consists of off-axis parabolic mirrors (OAPs) and lenses, with OAPs positioned exclusively upstream of the FPM. The deformable mirrors (DMs) are reflective elements: DM2 is placed in the pupil plane, while DM1 is positioned a certain distance upstream of DM2. For clarity, all components are shown as transmissive elements. (b)-(d) SPLC and 2DM-SPLC design results for EAST and LOT. They were designed to achieve a contrast ratio of $1\times10^{-7}$, prioritizing higher throughput and coverage angle (see Section 3.1). The left column illustrates the telescope aperture geometries, while the right column presents the corresponding designs for each optical element shown in (a). For the DMs, the final optimized surface height maps are displayed. The three SPLC binary amplitude masks are shown, with white indicating the transmissive regions. The camera images depict the distribution of the captured dark hole. A shared colorbar is used across all columns.

**Convex optimization for SP and LS.** The SP, FPM, and LS play complementary roles: (1) The SP is the main optimization variable and redistributes the pupil illumination to form a prescribed dark hole, such as a bowtie-shaped region, in the focal plane. (2) The FPM blocks the stellar core while transmitting the dark-hole region and is fixed by the chosen dark-hole geometry. (3) The LS suppresses diffracted starlight from the aperture edges and upstream masks, and is parameterized as an annulus with inner and outer diameters.

To preserve convexity, we perform a nested optimization: The LS diameters ($r_{0,\mathrm{LS}}$ and $r_{1,\mathrm{LS}}$) are scanned on a grid, and for each candidate LS, a linear convex subproblem is solved to optimize the SP. The final SPLC design is selected according to the trade-off between contrast and throughput:

$$\begin{aligned}
&\text{For each:} \quad r_{0,\mathrm{LS}},\ r_{1,\mathrm{LS}} \in [0, D], \quad r_{0,\mathrm{LS}} < r_{1,\mathrm{LS}} \\
&\text{Maximize:} \quad J = \mathrm{sum}\left[M_{\mathrm{SP}}\left(x_i, y_j\right)\right], \\
&\text{Subject to:}\ 0 \le M_{\mathrm{SP}}\left(x_i, y_j\right) \le 1,\ \left(x_i, y_j\right) \in M_{\mathrm{AP}}, \\
&\qquad -E_{\mathrm{Norm}}\sqrt{C_{\mathrm{aim}}} \le E_{\mathrm{Cam}} \le E_{\mathrm{Norm}}\sqrt{C_{\mathrm{aim}}}, \quad (5)
\end{aligned}$$

where $D$ is the telescope aperture diameter, $r_{0,\mathrm{LS}}$ and $r_{1,\mathrm{LS}}$ are the inner and outer diameters of the LS, and $C_{\mathrm{aim}}$ represents the target contrast. To improve robustness against vibration and drift, the SP gap width in the aperture constraint $\left(x_i, y_j\right) \in M_{\mathrm{AP}}$ is set larger than the telescope segment gaps. The SP transmission is optimized as a continuous variable between 0 and 1; because the objective maximizes throughput under the dark-hole constraint, the solution tends to be nearly binary without explicitly imposing a binarization constraint.

### 2.3 Optimal design of 2DM-SPLC

Compared with a pure SPLC, the 2DM-SPLC includes two DMs to provide additional degrees of freedom for complex-field control in the dark hole. The pupil-plane DM provides phase modulation, while the upstream non-pupil DM enables coupled phase and amplitude modulation through propagation, allowing the system to further suppress stellar intensity and correct residual aberrations.

We adopt the EFC algorithm [79] to compute both the preloaded DM patterns for static dark-hole generation and the dynamic commands for wavefront correction. Let $\Delta u_{q,t}$ be the voltage increment applied to actuator $q$ at iteration $t$, and let $a_q f_q$ denote the corresponding influence function (with gain $a_q$). After $k$ iterations, the accumulated DM surface heights can be written as

$$\phi_k = \sum_{t=1}^{k} \Delta\phi_t = \sum_{t=1}^{k} \sum_{q=1}^{N_{act}} \Delta u_{q,t} a_q f_q = \phi_{k-1} + \phi(\Delta u_k), \quad (6)$$

where $\Delta u_k$ collects the voltage updates for all actuators at iteration $k$, and $\phi(\Delta u_k)$ is the induced surface-height increment.

Based on the SPLC transfer operator in Eq. (3) and the DM surface-height representation in Eq. (6), and denoting Fresnel propagation over a distance $z_i$ by $\mathcal{F}r_{z_i}$, the coronagraphic field at iteration $k$ is

$$\begin{aligned} E_{C,k} &= T_C\left[E_{\text{in},k}\right] = T_C\left\{\exp\left(i\frac{4\pi\phi_{2,k}}{\lambda}\right)\cdot\mathcal{F}r_{z_2}\left\{\mathcal{F}r_{z_1}\{E_a\}\cdot\exp\left(i\frac{4\pi\phi_{1,k}}{\lambda}\right)\right\}\right\} \\ &= T_C\left\{\exp\left(i\frac{4\pi\left[\phi_{2,k-1}+\phi\left(\Delta u_{2,k}\right)\right]}{\lambda}\right)\cdot\mathcal{F}r_{z_2}\left\{\mathcal{F}r_{z_1}\{E_a\}\cdot\exp\left(i\frac{4\pi\left[\phi_{1,k-1}+\phi\left(\Delta u_{1,k}\right)\right]}{\lambda}\right)\right\}\right\} \end{aligned} \quad (7)$$

where $E_a$ is the complex field immediately before the telescope aperture, $\mathcal{F}r_{z_1}$ propagates this field from the telescope aperture plane to DM1, and $\mathcal{F}r_{z_2}$ propagates the field from DM1 to DM2, which is located at the pupil plane. $\phi_{1,k}$ and $\phi_{2,k}$ are the DM1 and DM2 surface heights. We linearize Eq. (7) by first-order Taylor expansion of the two DM phase factors, yielding the following relationship,

$$\begin{aligned} E_{C,k} &\approx E_{C,k-1} + i\frac{4\pi}{\lambda}T_C\left\{E_{\text{in},k-1}\cdot\phi\left(\Delta u_{2,k}\right)\right\} \\ &+ i\frac{4\pi}{\lambda}T_C\left\{\exp\left(i\frac{4\pi\phi_{2,k-1}}{\lambda}\right)\cdot\mathcal{F}r_{z_2}\left\{\mathcal{F}r_{z_1}\{E_a\}\cdot\exp\left(i\frac{4\pi\phi_{1,k-1}}{\lambda}\right)\cdot\phi\left(\Delta u_{1,k}\right)\right\}\right\} \end{aligned} \quad (8)$$

The two linearized perturbation terms in Eq. (8) can be written as the responses of two Jacobian matrices.

$$\begin{aligned} E_{C,k} &\approx E_{C,k-1} + G_1\Delta u_{1,k} + G_2\Delta u_{2,k} \\ &= E_{C,k-1} + G\Delta u_k \end{aligned} \quad (9)$$

where $G_1$ and $G_2$ are the Jacobian matrix (or response matrix) mapping the actuator-command increments of DM1 and DM2 to the complex electric-field variation in the dark hole, respectively. The combined Jacobian matrix is $G = [G_1, G_2]$, and $\Delta u_k$ is the concatenated command vector of $\Delta u_{1,k}$ and $\Delta u_{2,k}$. $G$ can be computed based on the Fourier optics model of the SPLC system. Thus, $G\Delta u_k$ represents the first-order change in the coronagraphic electric field produced jointly by the two DMs.

The EFC update at iteration $k$ is obtained by minimizing a quadratic cost function that penalizes both residual field and control effort:

$$\begin{aligned} &\min_{\Delta u_k} J_k = E_{C,k}^H E_{C,k} + \alpha\Delta u_k^T\Delta u_k \\ &s.t.\ \ E_{C,k} = E_{C,k-1} + G\Delta u_k \end{aligned} \quad (10)$$

where $\alpha$ is a Tikhonov regularization parameter. $[\cdot]^H$ and $[\cdot]^T$ denote the Hermitian transpose and transpose, respectively. Taking the derivative of the cost function $J_k$ with respect to $\Delta u_k$ and setting it to zero yields the standard EFC update

$$\Delta u_k = -(\mathrm{Re}[G^H G] + \alpha I)^{-1}\cdot\mathrm{Re}\left[G^H E_{C,k-1}\right] \quad (11)$$

where $\mathrm{Re}[\cdot]$ denotes the real part, and $I$ represents the identity matrix. After several iterations, $\Delta u_k$ converges to zero, and the dark-hole contrast stabilizes. The optimized commands for the two DMs, $u_{1,k}$ and $u_{2,k}$, are stored as the preloaded DM patterns for the 2DM-SPLC. The full 2DM-SPLC design is decomposed into three tractable steps: SP optimization, LS diameter search, and EFC-based computation of static DM preloads.

### 2.4 Wavefront error simulation and correction

For ground-based coronagraphs, wavefront errors mainly arise from time-varying atmospheric turbulence and quasi-static instrumental aberrations. Here, we focus on atmospheric phase errors, while quasi-static terms are included in the general residual error budget. Both the EAST and LOT telescopes are planned to be located at the Lenghu site in Qinghai Province [89], where the median long-exposure seeing is approximately 0.75" at 550 nm. Assuming Kolmogorov turbulence, the seeing $\varepsilon$ and the Fried parameter $r_0$ are related by [90]

$$\varepsilon = 0.98\frac{\lambda}{r_0}, \quad (12)$$

where $\varepsilon$ is the full width at half maximum (FWHM) of the long-exposure PSF. The corresponding wavelength scaling is

$$\varepsilon \propto \lambda^{-\frac{1}{5}}, \qquad r_0 \propto \lambda^{\frac{6}{5}}. \quad (13)$$

For a telescope of diameter $D$, the Kolmogorov phase variance $\sigma_\phi^2$ over the pupil is [91,92]

$$\sigma_\phi^2 \simeq 1.03\left(\frac{D}{r_0}\right)^{\frac{5}{3}}. \quad (14)$$

Atmospheric phase screens are synthesized by expanding the phase $\phi(\rho,\theta)$ into real-valued Zernike polynomials on the unit circular aperture. The modal coefficients follow the Kolmogorov spectrum, with the variance of modes of radial order (n) scaled as [91]

$$\mathrm{Var}(a_n) \propto (n+1)^{-\frac{11}{3}}. \quad (15)$$

A global normalization is then applied to all coefficients such that the root mean square (RMS) of $\phi(\rho,\theta)$ within the telescope pupil matches $\sigma_\phi$, derived from the selected seeing condition using Eq. (12)– (14).

The residual wavefront after telescope-side correction is modeled in three stages: (1) Piston is excluded, and tip/tilt is removed by a telescope-side tip/tilt mirror, accounting for approximately 87% of the Kolmogorov phase variance [93]. (2) The telescope-side low-order AO stage is then modeled by applying a 60% effective correction to the remaining Zernike components, mainly modes 4–11, with the uncorrected fraction representing wavefront-sensor noise, reconstruction errors, fitting errors, and other non-ideal effects. (3) The residual high-order aberrations are subsequently corrected by the two deformable mirrors in the 2DM-SPLC, which serve as a coronagraph-side extreme adaptive optics stage. For this upper-bound evaluation, a noise-free focal-plane sensing model and electric field conjugation control are used to suppress residual atmospheric speckles within the controllable spatial-frequency range. The resulting contrast therefore represents an upper-bound estimate of the 2DM-SPLC performance under Kolmogorov turbulence at the Lenghu site.

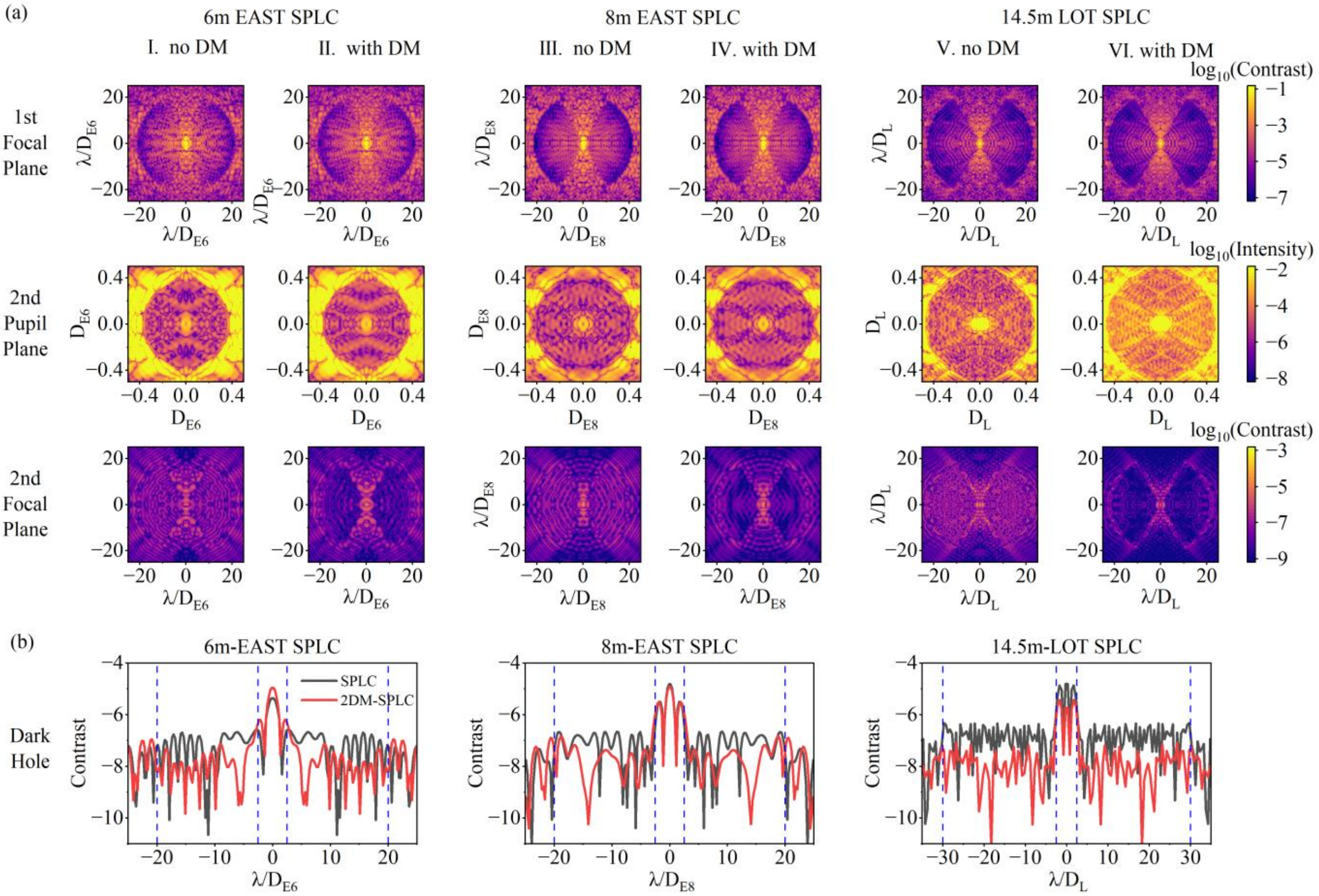

**Fig. 2. Optical simulation results of the SPLC and 2DM-SPLC.** (a) Electric field distributions of the SPLC and 2DM-SPLC for 6m-EAST, 8m-EAST, and 14.5m-LOT at the first focal plane, second pupil plane, and second focal plane. The focal plane panels display the contrast maps, while the pupil plane panels show the intensity distributions at the respective planes for a normalized uniform input wavefront. The colorbar is shared across all rows. (b) Comparison of the dark-hole contrast achieved with the SPLC and 2DM-SPLC. The blue dashed lines indicate the boundaries of the dark hole. All three panels share the same legend.

# 3. RESULTS

## 3.1 The Design of SPLC and 2DM-SPLC

We propose an SPLC design tailored for ground-based telescopes facing atmospheric turbulence. To balance practical feasibility with scientific requirements, we set the target contrast to $1 \times 10^{-7}$ , enabling larger dark holes and higher throughput. Our SPLC design is implemented in CVXPY [94] as a nested optimization problem, scanning LS diameter to maximize throughput under a target contrast. The key design parameters—bandwidth, IWA/ Outer Working Angle (OWA), field coverage, and contrast requirement for each telescope configuration—are summarized in Table 2.

**Table 2. The design parameters of SPLC**

| Factors | 6m-EAST | 8m-EAST | 14.5m-LOT |
|---|---|---|---|
| Bandwidth $\Delta\lambda/\lambda$ (%) | 10 | 10 | 10 |
| IWA ($\lambda/D$) | 2.5 | 2.5 | 2.5 |
| OWA ($\lambda/D$) | 20 | 20 | 30 |
| Field Coverage Angle (°) | 240 | 240 | 240 |
| Target Contrast | $1 \times 10^{-7}$ | $1 \times 10^{-7}$ | $1 \times 10^{-7}$ |

The final SPLC designs for EAST and LOT are shown in Fig. 1(b)-(d). Based on the grid search on LS diameter (see Fig. 4), the optimal throughput is achieved when the LS inner and outer diameters are set to 0.1 $D$ and 0.76 $D$ for 6m-EAST, 0.1$D$ and 0.86 $D$ for 8m-EAST, and 0.15 $D$ and 0.95 $D$ for 14.5m-LOT (Fig. 1(b-I, c-I, d-I)). Fig. 2(a-I, III, V) shows the SPLC light field in the first focal plane (FPM), second pupil plane (LS), and final focal plane. The resulting mean contrasts and throughputs are $7.52 \times 10^{-8}$ and 0.3001 for 6m-EAST, $8.20 \times 10^{-8}$ and 0.3203 for 8m-EAST; $7.85 \times 10^{-8}$ and 0.4746 for 14.5m-LOT.

After optimizing the DM surface shapes with the EFC algorithm, we obtain improved 2DM-SPLC designs. For EAST, we use two 34×34-actuator DMs and optimize the dark hole over 2.5–20 $\lambda/D$; for LOT, we adopt 50×50-actuator DMs and a wider dark hole of 2.5–30 $\lambda/D$. The final 2DM-SPLC configurations are shown in Fig. 1(b-II, c-II, d-II), where both DMs are preloaded with optimized surface profiles in the range of −20 to 20 nm. Fig. 2(a-II, IV, VI) presents the corresponding electric field distributions. With the two DMs included, the mean dark-hole contrast improves to $1.55 \times 10^{-8}$ for 6m-EAST, $2.68 \times 10^{-8}$ for 8m-EAST, and $1.29 \times 10^{-8}$ for 14.5m-LOT. As summarized in Fig. 2(b), the 2DM-SPLC consistently outperforms the SPLC alone, confirming that added pupil-plane phase control further deepens the dark hole.

## 3.2 Performance of 2DM-SPLC under Kolmogorov turbulence

Following the atmospheric-turbulence model described in Sec. 2.4, we evaluate the 2DM-SPLC performance under three representative Lenghu seeing conditions: 0.51″, 0.75″, and 1.03″ at 550 nm, corresponding to the 10%, median, and 75% of observation time [89]. These seeing values are scaled to the V (Hα, 656.3 nm), H (1.6 μm), and K (2.2 μm) bands using Kolmogorov laws, and the corresponding phase screens are generated with the Zernike-based model up to radial order 50 (1325 modes excluding piston). These aberrations are applied to the 6m-EAST, 8m-EAST, and 14.5m-LOT pupils.

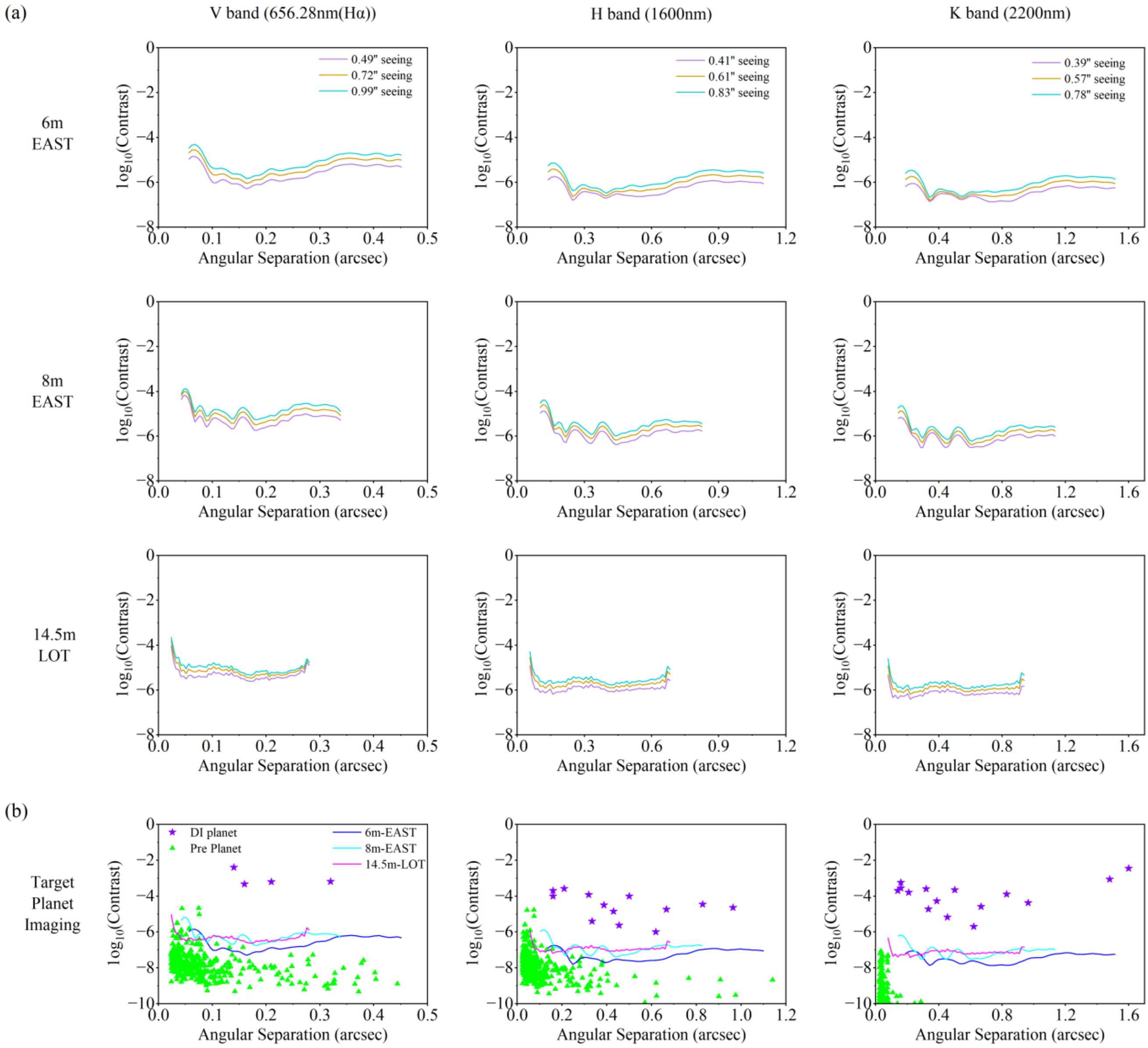

**Fig. 3. Performance of the 2DM-SPLC under wavefront errors.** (a) Dark-hole contrast of the 2DM-SPLC for different seeing conditions, bands, and apertures. Each row represents apertures of 6m-EAST, 8m-EAST, and 14.5m-LOT, respectively. Each column corresponds to the V, H, and K bands, with the same legend for each column. The gray dashed lines indicate the positions of the IWA and OWA of the 2DM-SPLC. (b) Dark-hole contrast of the 2DM-SPLC designed for 6m-EAST, 8m-EAST, and 14.5m-LOT in each band under good seeing conditions, along with data for the target planets. Purple pentagons indicate planets that have been directly imaged, while green triangles indicate catalogued non-directly imaged exoplanets whose contrasts are estimated in this work.

Before correction, the RMS phase errors are 1407 nm, 1940 nm, and 2528 nm for these three seeing conditions, corresponding to a contrast of approximately $10^{-3}$. After ideal tip/tilt correction, which removes approximately 87% of the Kolmogorov phase variance, the residual RMS values decrease to 507 nm, 889 nm, and 1158 nm. Accounting for wavefront-sensor measurement errors, the subsequent low-order AO stage is assumed to correct 60% of the residual aberrations, mainly Zernike modes 4–11, reducing the RMS values to 366 nm, 613 nm, and 799 nm, respectively. The remaining residual wavefront errors are then corrected by the coronagraph-side 2-DM ExAO system in the 2DM-SPLC. The resulting dark-hole contrasts are shown in Fig. 3(a). As expected, the mean dark-hole contrast degrades as the seeing condition becomes worse.

For the favorable seeing case (0.51″ at 550 nm), the low-order AO residual RMS is 366 nm. We use this common input to compare the ExAO correction performance of the three telescope/DM configurations. The 2DM-SPLC reduces the input RMS to 169 nm, 219 nm, and 303 nm for the 6m-EAST, 8m-EAST, and 14.5m-LOT, respectively. Although the LOT has more actuators, its much larger aperture results in coarser actuator sampling in units of the pupil diameter, leaving a larger fraction of high-spatial-frequency aberrations uncorrected. This indicates that larger segmented apertures require higher actuator densities to achieve comparable ExAO correction performance.

Although the atmospheric residuals degrade the contrast relative to the nominal static design, the corrected 2DM-SPLC maintains raw contrasts of order $10^{-6}$ – $10^{-7}$ under favorable seeing conditions. Therefore, when designing SPLCs for ground-based large segmented telescopes, the target contrast should be chosen close to the contrast achievable after atmospheric correction, which helps maximize the overall SPLC performance. Across the three bands, longer wavelengths yield smaller effective seeing and thus more favorable high-contrast performance. The angular scale of the dark hole follows $\lambda/D$: increasing the wavelength expands the IWA and OWA in arcseconds, whereas increasing the telescope aperture reduces them. However, the useful OWA is ultimately limited by the spatial-frequency range of the DMs. Although the SPLC can be optimized for a larger OWA, extending the dark hole to larger angular separations requires stronger high-order wavefront correction and is therefore constrained by the finite DM actuator density. Table 3 summarizes the IWA, OWA, and mean raw contrast of the 2DM-SPLC under favorable seeing conditions for each band.

**Table 3. Performance of the 2DM-SPLC in different bands under good seeing**

**I. V band**

| Telescope | IWA | OWA | Raw Contrast |
|---|---|---|---|
| 6m-EAST | 0.0564″ | 0.4512″ | $2.51 \times 10^{-6}$ |
| 8m-EAST | 0.0423″ | 0.3384″ | $5.72 \times 10^{-6}$ |
| 14.5m-LOT | 0.0233″ | 0.2801″ | $4.68 \times 10^{-6}$ |

**II. H band**

| Telescope | IWA | OWA | Raw Contrast |
|---|---|---|---|
| 6m-EAST | 0.1375″ | 1.1001″ | $5.33 \times 10^{-7}$ |
| 8m-EAST | 0.1031″ | 0.8251″ | $1.29 \times 10^{-6}$ |
| 14.5m-LOT | 0.0569″ | 0.6828″ | $1.19 \times 10^{-6}$ |

**III. K band**

| Telescope | IWA | OWA | Raw Contrast |
|---|---|---|---|
| 6m-EAST | 0.1891″ | 1.5126″ | $3.36 \times 10^{-7}$ |
| 8m-EAST | 0.1419″ | 1.1345″ | $8.20 \times 10^{-7}$ |
| 14.5m-LOT | 0.0782″ | 0.9389″ | $7.29 \times 10^{-7}$ |

Fig. 3(b) compares the post-processed 2DM-SPLC contrast curves with the angular separations and planet-to-star contrasts of known directly imaged exoplanets, whose parameters are listed in Table 4. A post-processing gain factor of 10 is applied to the raw contrast curves to provide a representative estimate of the achievable detection limits. The results indicate that both EAST and LOT could support follow-up observations of these known systems and may help search for additional companions in similar environments. To assess the broader discovery potential, we further estimate the contrasts of 366 catalogued exoplanets without direct-imaging detections and overplot them in Fig. 3(b). The model includes both reflected light, based on wavelength- and planet-type-dependent geometric albedos [95], and thermal emission under radiative equilibrium, with the planet-to-star thermal contrast computed from the Planck function [96,97]. The planetary and stellar parameters are taken from the NASA Exoplanet Archive [98]. A planet is regarded as potentially accessible when its predicted contrast at the corresponding angular separation is above the post-processed instrumental contrast curve.

Most of the 366 planets lie at small angular separations, so their accessibility is mainly governed by the trade-off between IWA and achievable contrast. In the V-band, the 2DM-SPLC reaches the smallest angular separations but with relatively modest contrast, yielding about 25 potentially accessible planets. In the H-band, the deeper contrast compensates for the larger inner working angle, giving approximately 24 accessible objects. The K-band provides a smaller accessible subset because reflected light is weak and the thermal emission of most mature cool planets peaks at longer wavelengths; the accessible targets are therefore mainly young and massive giant planets with substantial residual heat. These results suggest that large segmented telescopes equipped with 2DM-SPLC systems could expand the accessible target space for future direct-imaging observations.

**Table 4. Parameters of discovered planets through direct imaging**

| Planet Name | Separation | V-band Contrast | H-band Contrast | K-band Contrast |
|---|---|---|---|---|
| PDS 70 b[27,99,100] | 0.16″ | $4.7 \times 10^{-4}$ | $2 \times 10^{-4}$ | $5.8 \times 10^{-4}$ |
| PDS 70 c[26,99,100] | 0.21″ | $6.2 \times 10^{-4}$ | $2.6 \times 10^{-4}$ | $1.6 \times 10^{-4}$ |
| WISPIT 2 b[101,102] | 0.32″ | $6.5 \times 10^{-4}$ | $1.2 \times 10^{-4}$ | $2.5 \times 10^{-4}$ |
| 2MJ1612 b[103,104] | 0.14″ | $4 \times 10^{-3}$ | / | $2 \times 10^{-4}$ |
| AF Lep b[105] | 0.33″ | / | $3.9 \times 10^{-6}$ | $1.9 \times 10^{-5}$ |
| β Pic b [106] | 0.50″ | / | $1 \times 10^{-4}$ | $2.2 \times 10^{-4}$ |
| HD 95086 b [107] | 0.62″ | / | $1 \times 10^{-6}$ | $2 \times 10^{-6}$ |
| HD 135344 Ab [108] | 0.16″ | / | $9.9 \times 10^{-5}$ | $2.8 \times 10^{-4}$ |
| HD 143811 AB b [109] | 0.43″ | / | $1.4 \times 10^{-5}$ | / |
| HIP 65426 b [24] | 0.83″ | / | $3.5 \times 10^{-5}$ | $1.3 \times 10^{-4}$ |
| HR 8799 c [110] | 0.96″ | / | $2.3 \times 10^{-5}$ | $4.2 \times 10^{-5}$ |
| HR 8799 d [110] | 0.67″ | / | $1.8 \times 10^{-5}$ | $2.6 \times 10^{-5}$ |
| HR 8799 e [110] | 0.39″ | / | $3.2 \times 10^{-5}$ | $5.3 \times 10^{-5}$ |
| WISPIT 1 b [111] | 1.48″ | | $3.8 \times 10^{-4}$ | $8.6 \times 10^{-4}$ |
| YSES-1 b [112] | 1.60″ | / | $2.2 \times 10^{-3}$ | $3.5 \times 10^{-3}$ |
| 51 Eri b [113] | 0.45″ | / | $2.3 \times 10^{-6}$ | $6.7 \times 10^{-6}$ |

## 4. DISCUSSION

### 4.1 Relationship between contrast and throughput

Contrast and throughput are two key metrics for evaluating coronagraph performance. For SPLC designs, a more stringent target contrast generally requires stronger pupil apodization, which reduces throughput. As shown in Fig. 5, the throughput is only weakly affected at moderate contrast requirements, but decreases rapidly when the target contrast becomes too aggressive, making the design impractical. Among the three configurations, LOT maintains the highest throughput at a given target contrast, and its larger aperture further increases the photon-collecting capability.

### 4.2 Shape of the Lyot Stop

The LS suppresses diffracted stellar light from the telescope aperture, SP, and FPM. In this work, the LS is parameterized as an annulus, with its inner and outer diameters scanned on a grid to maximize throughput at a fixed target contrast. To examine the influence of LS geometry, we adopt a stringent design case with a target contrast of $1 \times 10^{-10}$ and a coverage angle of 180°, while keeping the other parameters consistent with Table 2. Fig. 4 shows the optimized throughput as a function of LS inner and outer diameters for 6m-EAST, 8m-EAST, and 14.5m-LOT. For EAST, most diffracted starlight lies outside the geometric pupil, so a ring-shaped LS without central obscuration achieves the required dark-hole contrast while preserving high throughput. For LOT, however, intra-pupil diffracted light becomes more significant, and the optimal LS includes a central obscuration. This modestly reduces throughput but is required to satisfy the contrast constraint.

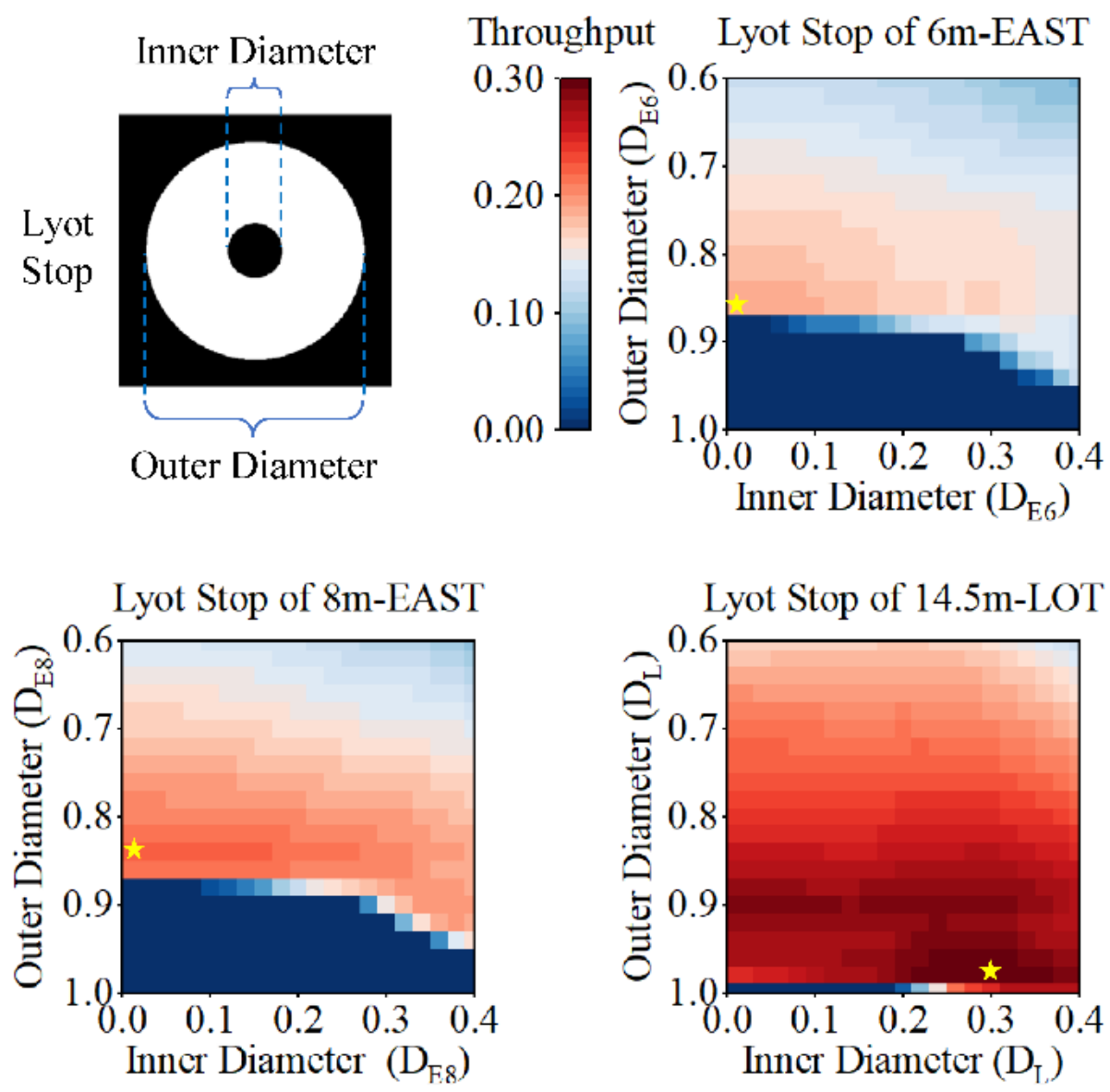

**Fig. 4. Dependence of SPLC throughput on the inner and outer diameters of the ring-shaped LS.** The upper-left panel illustrates the two parameters of the ring-shaped LS, and the remaining panels show the optimized throughput for the 6m-EAST, 8m-EAST, and 14.5m-LOT, respectively. Each pixel represents the throughput of the optimized SPLC corresponding to specific inner and outer diameters of the LS. The yellow pentagram marks the design with the maximum throughput. All three panels use the same colorbar.

More generally, LS geometry can be tailored to specific science cases. With a ring-shaped FPM, scattered light is concentrated mainly at low spatial frequencies, making a ring-shaped LS effective. For a bowtie-shaped FPM, scattered light is mainly distributed along the bowtie directions, and a bowtie-shaped LS can block these regions selectively. Fig. 5 compares the contrast–throughput curves for ring-shaped and bowtie-shaped LS designs. The bowtie-shaped LS achieves higher contrast by blocking more scattered light, but at moderate contrast requirements it also reduces throughput more strongly. Thus, when ultra-high contrast is not required, a ring-shaped LS provides a good compromise, suppressing the dominant scattered light while maintaining relatively high throughput.

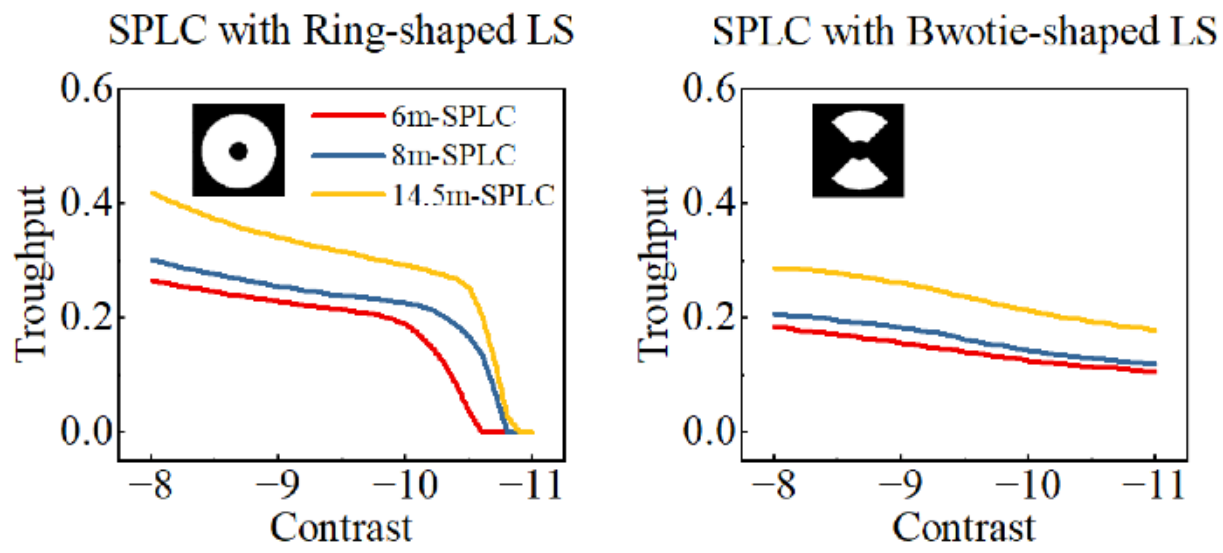

**Fig. 5. Contrast–throughput trade-off under ring-shaped and bowtie-shaped LS configurations.** The panels compare the contrast and throughput of the SPLC designs for the 6m-EAST, 8m-EAST, and 14.5m-LOT telescopes, each evaluated with two different LS shapes. A throughput value of zero indicates that the SPLC design is invalid. All panels share the same legend.

### 4.3 Bandwidth and dark-hole geometry

In the SPLC design, several target parameters must be specified, and their influence on performance should be understood to match different science goals. To isolate these effects, we fix the target contrast at $1 \times 10^{-9}$, use the baseline parameters in Table 2, and vary one parameter at a time, with throughput adopted as the performance metric.

The observing band is set by the expected planet spectrum. Near-infrared bands are generally preferred for exoplanet imaging because planets are brighter and atmospheric seeing is more favorable at longer wavelengths. Since the dark hole is defined in units of $\lambda/D$, changing the central wavelength mainly rescales its physical size and does not alter the SPLC pattern itself. However, bandwidth directly affects the optimization because the contrast constraint must be satisfied at multiple wavelengths. As shown in Fig. 6(a), the throughput drops noticeably as bandwidth increases. We also find that more than 5 wavelength samples provide nearly converged solutions.

The dark hole is characterized by the coverage angle, IWA, and OWA. The coverage angle sets the azimuthal extent of the dark region; increasing it improves field coverage and observational robustness but reduces throughput, as shown in Fig. 6(b). A smaller IWA makes starlight suppression more difficult and degrades SPLC performance, as seen in Fig. 6(c). By contrast, the OWA has a weaker impact on the SPLC solution (Fig. 6(d)): in principle it can be extended to larger angular separations, but in practice the useful dark-hole size is limited by the AO system's ability to correct high-spatial-frequency aberrations, as discussed in Sec. 3.2.

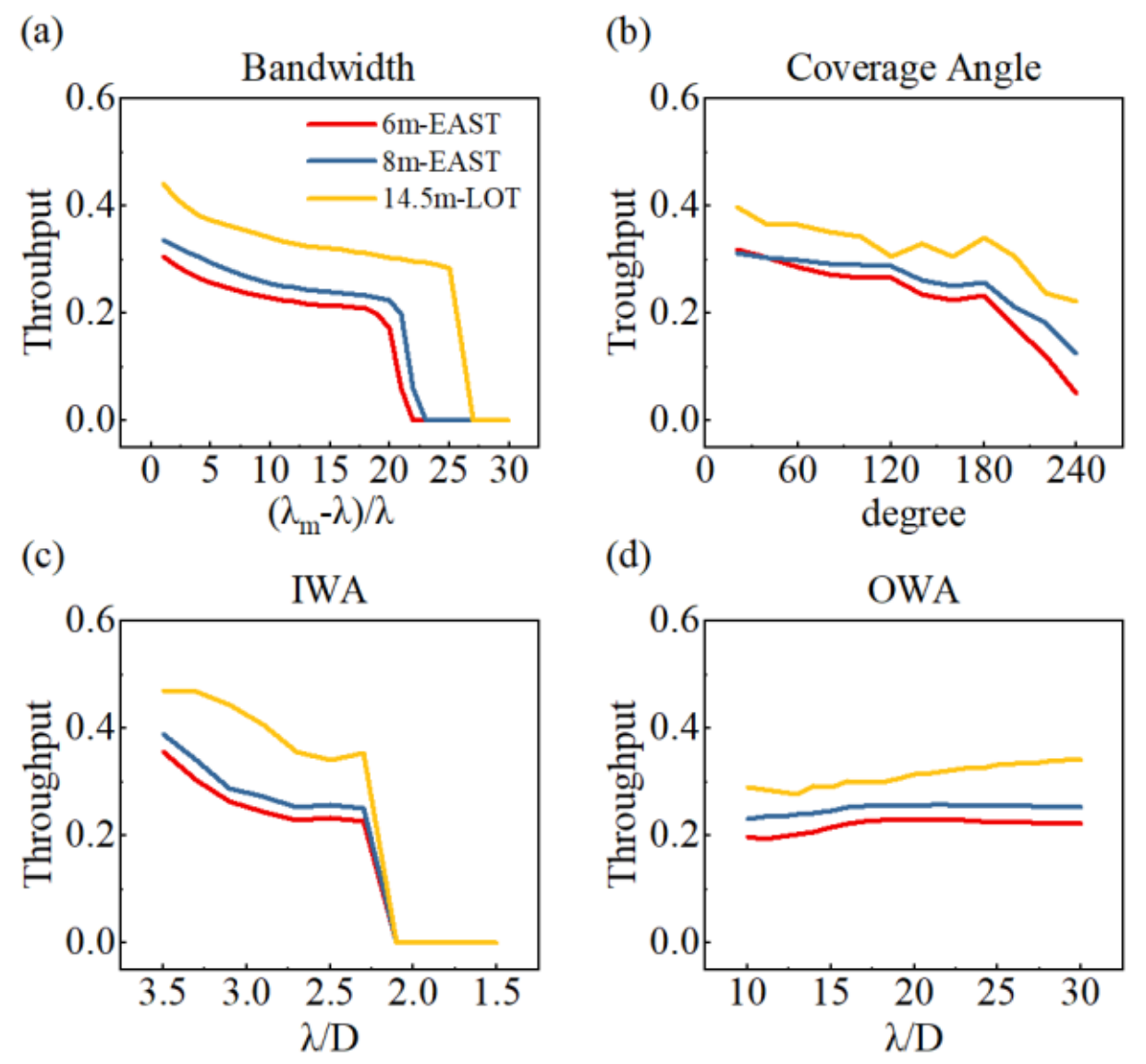

**Fig. 6. Dependence of SPLC performance on bandwidth and dark-hole geometry.** (a)–(d) show the throughput variations with bandwidth, coverage angle, IWA, and OWA, respectively. All panels share the same legend.

## 5. CONCLUSIONS

In this study, we developed a 2DM-SPLC design framework tailored to the segmented pupils of the planned 6/8 m EAST and 14.5 m LOT. By co-optimizing the SP and LS, the proposed designs achieve throughputs of approximately 0.3–0.5 at an IWA of 2.5 $\lambda/D$ for a target contrast of $1 \times 10^{-7}$ , providing a practical balance among diffraction suppression, IWA, and throughput. Under representative Lenghu turbulence conditions, the idealized ExAO simulations, which neglect the temporal evolution of atmospheric turbulence, yield raw contrasts on the order of $10^{-6} - 10^{-7}$ in the V, H, and K bands, representing an upper-bound estimate of the system performance. The designs remain sensitive to a range of known directly imaged planets and selected non-directly imaged candidates. In addition, atmospheric turbulence remains the dominant factor limiting the ultimate dark-hole contrast for ground-based systems. The contrast floor is dictated by the interplay between actuator density and residual turbulence strength. Accordingly, higher DM actuator density and improved wavefront-control accuracy would be required to reduce the residual speckle floor and more fully exploit the diffraction-suppression capability of the optimized SPLC on large segmented telescopes.

Several aspects of this work merit further refinement and extension. First, future work should incorporate realistic closed-loop FPSC operation, including the temporal evolution of atmospheric turbulence, finite-bandwidth wavefront sensing, detector noise, model inaccuracies, and sampling constraints, to obtain more engineering-realistic estimates of achievable contrast and stability. Second, the SPLC designs do not yet systematically include manufacturing tolerances, coating and surface-roughness errors, and non–common-path aberrations. Third, our results indicate that high-spatial-frequency residual aberrations remain significant for larger apertures, suggesting that future instruments should explore DMs with higher actuator counts or multi-conjugate DM architectures to better control broadband phase and amplitude errors. Finally, extending the proposed 2DM-SPLC design and parameter-sensitivity framework to other segmented or centrally obscured telescopes, and integrating it with integral-field spectroscopy, polarimetric imaging, and advanced post-processing algorithms, could provide a useful pathway toward future ground-based direct imaging and characterization of exoplanets and circumstellar disks.

## DECLARATION OF INTERESTS

The authors declare no competing interests.